\documentclass[prd,twocolumn,floatfix,amsmath,nofootinbib,amssymb,floatfix]{revtex4}
\usepackage{graphicx,color,dcolumn,booktabs,bm}
\usepackage{longtable,lscape}
\usepackage{txfonts}
\usepackage{overpic}
\usepackage{amssymb}
\usepackage{indentfirst}
\usepackage{feynmf}   
\usepackage{slashed}  
\usepackage{cases}
\usepackage{color}
\usepackage{multirow}
\usepackage{epstopdf}
\usepackage{enumerate}
\usepackage{graphicx,color,dcolumn,booktabs,bm}
\usepackage[colorlinks, citecolor=blue,anchorcolor=red,menucolor=red, linkcolor=red,filecolor=red,urlcolor=blue,frenchlinks=red]{hyperref}

\begin{document}

\title{The role of the $\omega(4S)$ and $\omega(3D)$ states in mediating the $e^+e^-\to \omega\eta$ and $\omega\pi^0\pi^0$ processes}
\author{Qin-Song Zhou$^{1,2}$}\email{zhouqs13@lzu.edu.cn}
\author{Jun-Zhang Wang$^{1,2}$}\email{wangjzh2012@lzu.edu.cn}
\author{Xiang Liu$^{1,2,3,4}$}\email{xiangliu@lzu.edu.cn}
\affiliation{$^1$School of Physical Science and Technology, Lanzhou University, Lanzhou 730000, China\\
$^2$Research Center for Hadron and CSR Physics, Lanzhou University $\&$ Institute of Modern Physics of CAS, Lanzhou 730000, China\\
$^3$Lanzhou Center for Theoretical Physics, Key Laboratory of Theoretical Physics of Gansu Province,
and Frontiers Science Center for Rare Isotopes, Lanzhou University, Lanzhou 730000, China\\
$^4$Joint Research Center for Physics, Lanzhou University and Qinghai Normal University, Xining 810000, China}

\date{\today}

\begin{abstract}

The $e^+e^-\to \omega\eta$ and $e^+e^-\to\omega\pi^0\pi^0$ processes are ideal platforms to search for higher $\omega$ states. Focusing on the observations of two enhancement structures around 2.2 GeV existing in $e^+e^-\to \omega\eta$ and $e^+e^-\to\omega\pi^0\pi^0$ at BESIII, we analyze how 
the $\omega(4S)$ and $\omega(3D)$ states play the role in the $e^+e^-\to \omega\eta$ and $e^+e^-\to\omega\pi^0\pi^0$ processes. The present study is supported by theoretical $\omega$ mesonic spectroscopy. 
For reproducing the data of the cross sections of $e^+e^-\to\omega\eta$ and $\omega\pi^0\pi^0$, the intermediate $\omega(4S)$ and $\omega(3D)$ should be introduced, which indicates that the enhancement structures around 2.2 GeV existing in $e^+e^-\to\omega\eta$ and $\omega\pi^0\pi^0$ contain the $\omega(4S)$ and $\omega(3D)$ signals. 
Nonetheless, in the process $e^+e^-\to\omega\eta$, the $\omega(4S)$ plays a dominant role, while the $\omega(4S)$ and $\omega(3D)$ have similar sizable contributions in the process of $e^+e^-\to\omega\pi^0\pi^0$, which leads to a difference in the line shape of enhancement structure in the cross sections under the interference effect.
Thus, we find a solution to alleviate the puzzling difference of resonance parameter of two reported enhancement structures around 2.2 GeV existing in $e^+e^-\to \omega\eta$ and $e^+e^-\to\omega\pi^0\pi^0$ at BESIII. 
The present study provides valuable information to construct $\omega$ meson family, which can be accessible at future experiment like BESIII. 
\end{abstract}

\maketitle

\section{Introduction}\label{sec1}

Studying hadron spectroscopy may provide important hint to understand the non-perturbative behavior of strong interaction. With the enhancement of precision in particle detection, more and more new hadronic states were observed in the past decades. These surprises often brought us delight. Until now, constructing light flavor meson family with higher states has become an interesting research issue, especially with the accumulation of the data of $e^+e^-$ annihilation into light hadrons at $\sqrt{s}\sim 2$ GeV \cite{BaBar:2019kds,BESIII:2020xmw,BESIII:2021bjn,BESIII:2020kpr,BESIII:2021uni,BESIII:2021lho}.

Recently, the BESIII Collaboration reported the measurement of the cross sections of the $e^+e^-\to \omega\eta$ \cite{BESIII:2020xmw} and $e^+e^-\to \omega\pi^0\pi^0$ \cite{BESIII:2021uni} at center of mass energies from $2$ GeV to 3.08 GeV.  Due to the constraint from symmetry, $e^+e^-\to \omega\eta$ \cite{BESIII:2020xmw} and $e^+e^-\to \omega\pi^0\pi^0$ \cite{BESIII:2021uni} are suitable processes to search for $\omega$ mesons. 
Here, two enhancement structures around 2.2 GeV were observed, which has resonance parameters listed as below
$M_1=2222\pm7\pm2,\,{\rm MeV}$, 
$\Gamma_1=59\pm30\pm6,\,{\rm MeV}$,
$M_2=2179\pm21\pm3,\,{\rm MeV}$, and 
$\Gamma_2=89\pm28\pm5.\,{\rm MeV}$.
Obviously, the determined resonance parameters ($M_1$, $\Gamma_1$) and ($M_2$, $\Gamma_2$) by BESIII via two processes $e^+e^-\to \omega\eta$ \cite{BESIII:2020xmw} and $e^+e^-\to \omega\pi^0\pi^0$ \cite{BESIII:2021uni}, respectively, are different. 

\begin{figure}[!htb]
  \centering
  \begin{tabular}{c}
  \includegraphics[width=200pt]{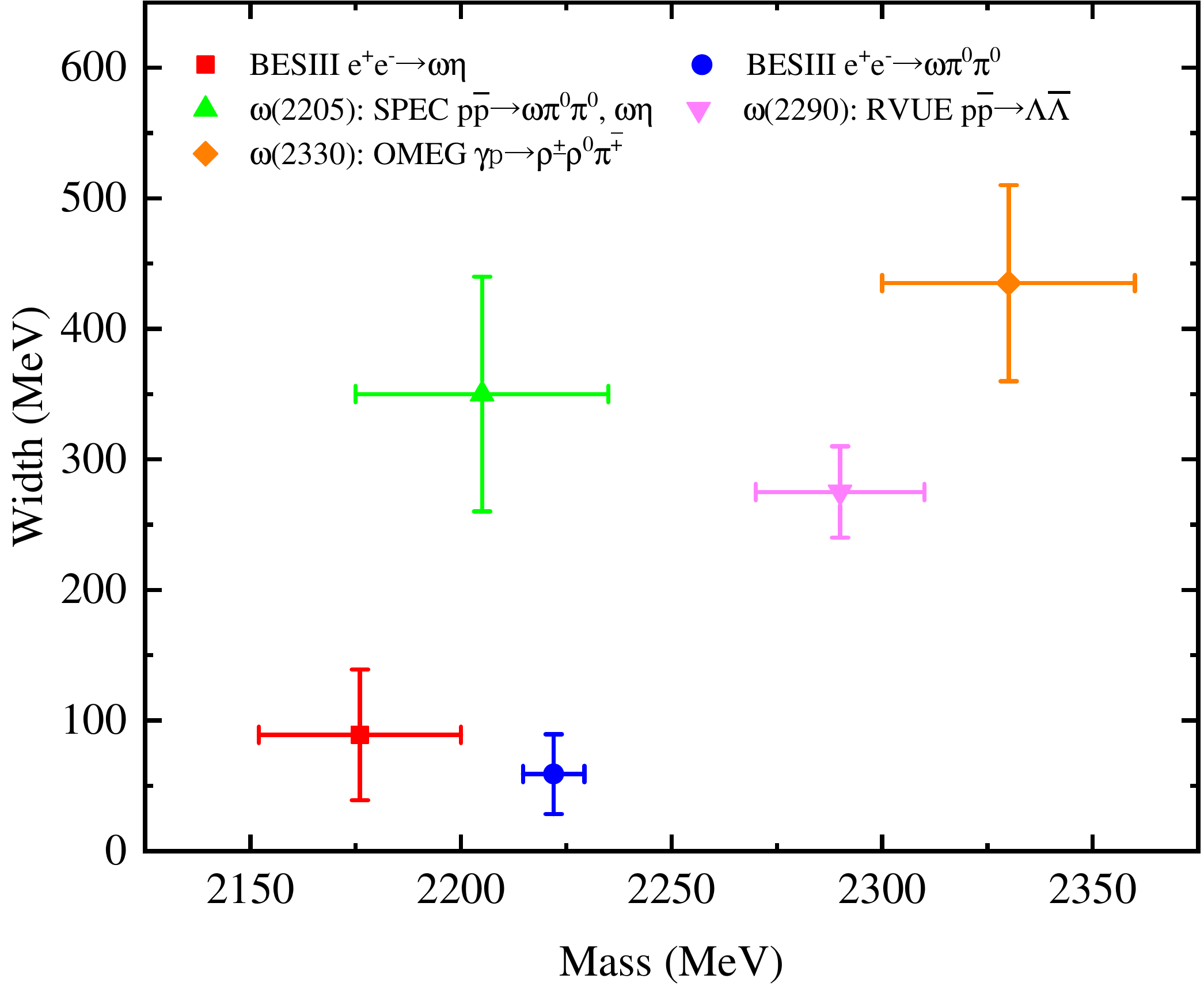}
  \end{tabular}
  \caption{A comparison of resonance parameters of these reported $\omega$ states with masses around 2.2 GeV \cite{BESIII:2020xmw,BESIII:2021uni,ParticleDataGroup:2020ssz}.}\label{RP}
\end{figure}

When looking at particle listings collected by Particle Data Group (PDG) \cite{ParticleDataGroup:2020ssz}, three $\omega$ states, the $\omega(2205)$, $\omega(2290)$, $\omega(2330)$, as further state was presented here. In Ref. \cite{Pang:2019ovr}, Lanzhou group studied the possibility of the $\omega(2290)$ and $\omega(2330)$ as $\omega(4^3S_1)=\omega(4S)$ state, and the $\omega(2205)$ as $\omega(3^3D_1)=\omega(3D)$ state. But, there still exists difference of theoretical result and experimental data of total width when making the above assignment \cite{Pang:2019ovr}. A main reason is that the $\omega(2205)$, $\omega(2290)$, $\omega(2330)$ states were not established in experiment. Their resonance parameters can only be as reference for constructing the $\omega(4S)$ and $\omega(3D)$. 
Additionally, when putting these reported $\omega$ states in experiments together, we can obviously find the difference of their resonance parameters as shown in Fig. \ref{RP}, where five $\omega$ states are accumulated in the same energy range, which is puzzling for us. This messy situation should be clarified by further theoretical and experimental effort. 

Focusing on the present observations of $\omega$ states at BESIII, we propose that it is a good chance to identify the $\omega(4S)$ and $\omega(3D)$ contributions through $e^+e^-\to \omega\eta$ \cite{BESIII:2020xmw} and $e^+e^-\to \omega\pi^0\pi^0$ \cite{BESIII:2021uni} since $e^+e^-$ collision is high precision experiment, which is different from the reaction processes of reporting the $\omega(2205)$, $\omega(2290)$, $\omega(2330)$.

In this work, we analyze the $e^+e^-\to \omega\eta$ and $\omega\pi^0\pi^0$ processes with the help of the theoretical input of $\omega$ mesonic spectroscopy. Our results show that the enhancement structures around 2.2 GeV existing in $e^+e^-\to \omega\eta$ \cite{BESIII:2020xmw} and $\omega\pi^0\pi^0$ \cite{BESIII:2021uni} can be due to the $\omega(4S)$ and $\omega(3D)$ contributions. 
Nonetheless, in process $e^+e^-\to\omega\eta$, the $\omega(4S)$ plays a dominant role, while the $\omega(4S)$ and $\omega(3D)$ have similar sizable contributions in the process $e^+e^-\to\omega\pi^0\pi^0$.
Obviously, this scheme makes us easily understand the puzzling difference of the measured resonance parameters of two resonance structures existing in $e^+e^-\to \omega\eta$ \cite{BESIII:2020xmw} and $e^+e^-\to \omega\pi^0\pi^0$ \cite{BESIII:2021uni} reported by BESIII. The present study may provide valuable information to construct the $\omega$ meson family. With accumulation of high precision data at future experiment, we also suggest our experiment colleague to test such scenario proposed in this work.

The paper is organized as follows. After the Introduction, we illustrate how to depict the cross sections of the $e^+e^-\to \omega\eta$ and $\omega\pi^0\pi^0$ processes with the theoretical support on mass spectrum, and decay behaviors of the $\omega$ meson family around 2.2 GeV in Sec. \ref{sec2}. 
In Sec. \ref{sec3}, we perform a fit on the experimental data of the Bonn cross sections of  $e^+e^-\to \omega\eta$ and $\omega\pi^0\pi^0$ measured by the BESIII Collaboration to decipher the contributions of excited $\omega(4S)$ and $\omega(3D)$, which can help us understand the nature of the enhancement structures around 2.2 GeV observed in these two processes. 
Finally, this work ends with a discussion and conclusion in Sec. \ref{sec4}.

\section{Depicting the cross sections of the $e^+e^-\to \omega\eta$ and $\omega\pi^0\pi^0$ processes}\label{sec2}

BESIII measured the cross sections of the $e^+e^-\to \omega\eta$ \cite{BESIII:2020xmw} and $\omega\pi^0\pi^0$ \cite{BESIII:2021uni} processes at $\sqrt{s}=2\sim 3.08$ GeV, by which the event accumulation around 2.2 GeV can be found. For depicting these enhancement structures around 2.2 GeV, in this work we consider the contribution of higher $\omega$ mesonic states. Thus, theoretical knowledge of $\omega$ mesons around 2.2 GeV is helpful to carry out such study.

\subsection{Theoretical $\omega$ mesonic states around 2.2 GeV}\label{sec2A}

\begin{figure}[!htb]
  \centering
  \begin{tabular}{c}
  \includegraphics[width=220pt]{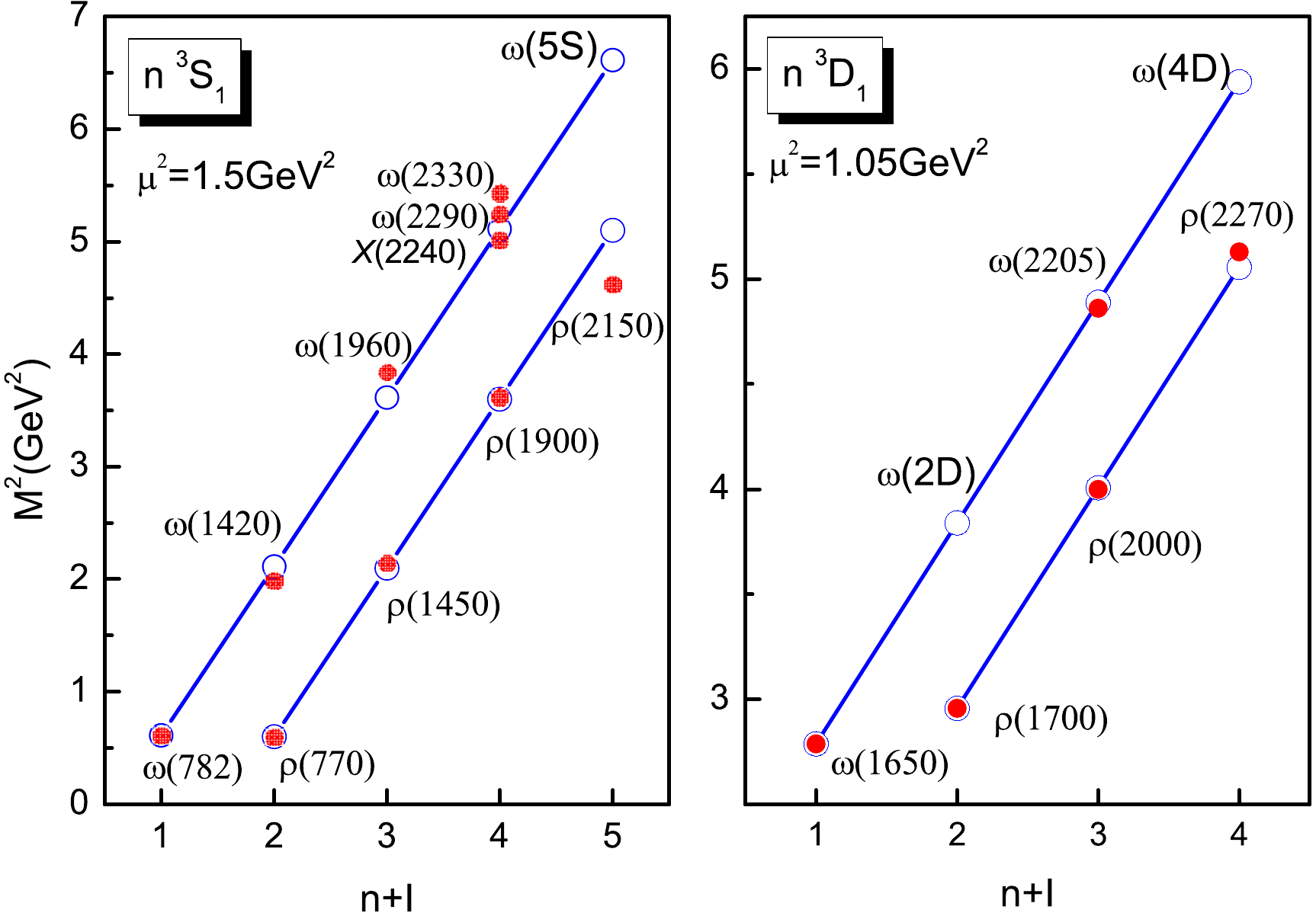}\\
  \end{tabular}
  \caption{The analysis of Regge trajectory for these reported $\omega$ states from PDG \cite{ParticleDataGroup:2020ssz}, and the comparison with the $\rho$ mesonic states \cite{Pang:2019ovr}. In general, the masses and radial quantum numbers of the light mesons in the same meson family should satisfy the relation $M^2=M_0^2+(n-1)\mu^2$ \cite{Chew:1962eu,Anisovich:2000kxa}, where $M_0$ denotes the mass of the ground state, $\mu^2$ is the trajectory slope and $n$ is the radial quantum number. In the diagram, $I$ denotes the isospin of the discussed mesons. The open circle and solid point correspond to the theoretical and experimental values, respectively.} \label{RT}
\end{figure}

Among these $\omega$ meson states collected in the Particle Data Group (PDG) \cite{ParticleDataGroup:2020ssz}, the $\omega(782)$ is well established as the ground state of S-wave $\omega$ meson. And then, the $\omega(1420)$ and $\omega(1960)$ are assigned as the first and the second radial excitations of the $\omega(782)$, respectively. As the ground state of D-wave $\omega$ meson, the $\omega(1650)$ is established (see Refs. \cite{Barnes:1996ff,Ebert:2005ha,Ebert:2009ub,Wang:2012wa,Pang:2019ovr} for more details). 
{Here, for illustrating the assignment to these observed $\omega$ states, we present the Regge trajectories for the $S$-wave and $D$-wave $\omega$ states, and make the comparison with the corresponding $\rho$ mesonic states, which are shown in Fig. \ref{RT}, by which the readers can learn the possible assignment \cite{Pang:2019ovr} to these observed $\omega$ states and their $\rho$ partners.}

\begin{figure}[!htb]
  \centering
  \begin{tabular}{c}
  \includegraphics[width=220pt]{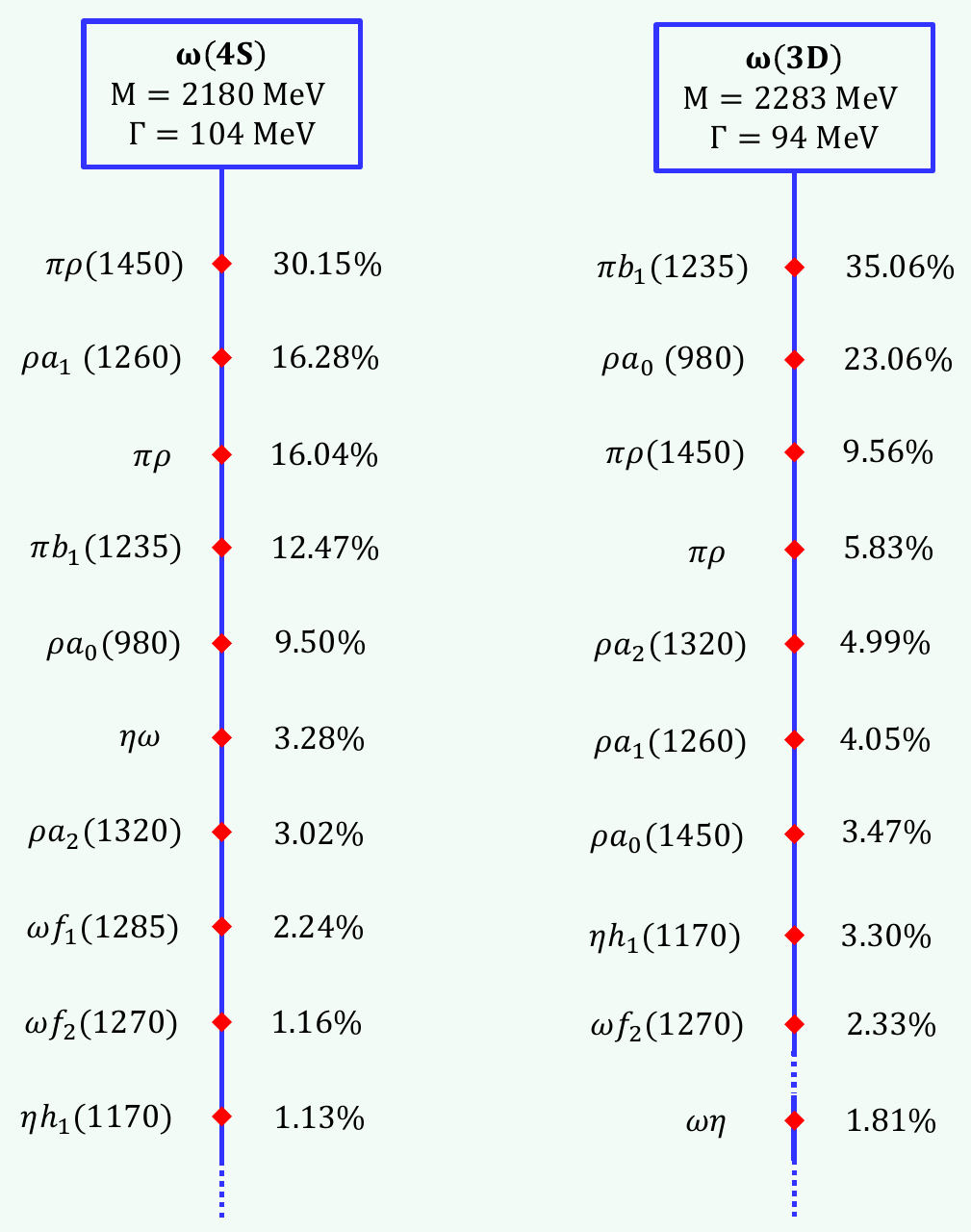}\\
  \end{tabular}
  \caption{The mass, total width, and the branching ratios of the OZI-allowed strong decays of the $\omega(4S)$ and $\omega(3D)$ states.}\label{OZIdecays}
\end{figure}

According to the theoretical study on the mass spectrum and decay behaviors of $\omega$ mesonic states \cite{Pang:2019ovr,Wang:2021gle}, the $\omega(4S)$ and $\omega(3D)$ were predicted with mass around 2.2 GeV. By an unquenched potential model, Lanzhou group studied the mass spectrum and the Okubo-Zweig-Iizuka (OZI)-allowed two-body strong decays of light flavor vector mesons \cite{Wang:2021gle}. 
Under this theoretical framework, the OZI-allowed two-body strong decays of the  $\omega(4S)$ and $\omega(3D)$ can be obtained. 
In Fig. \ref{OZIdecays}, we collect these theoretical results, which are applied to the following investigation for the cross section of the $e^+e^-\to \omega(4S)/\omega(3D) \to \omega\eta$ and $\omega\pi^0\pi^0$ processes.

\subsection{The $e^+e^-\to \omega\eta$ process}\label{sec2B}

\begin{figure}[!htb]
  \centering
  \begin{tabular}{cc}
  \includegraphics[width=110pt]{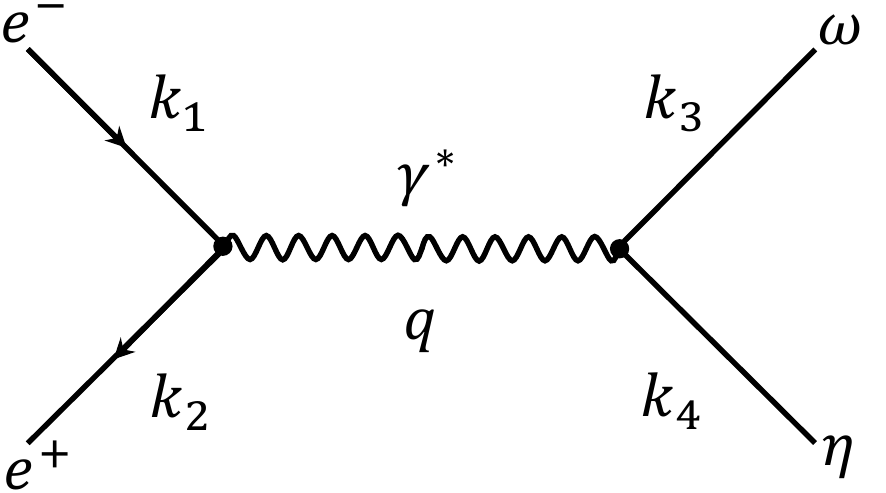}&\includegraphics[width=110pt]{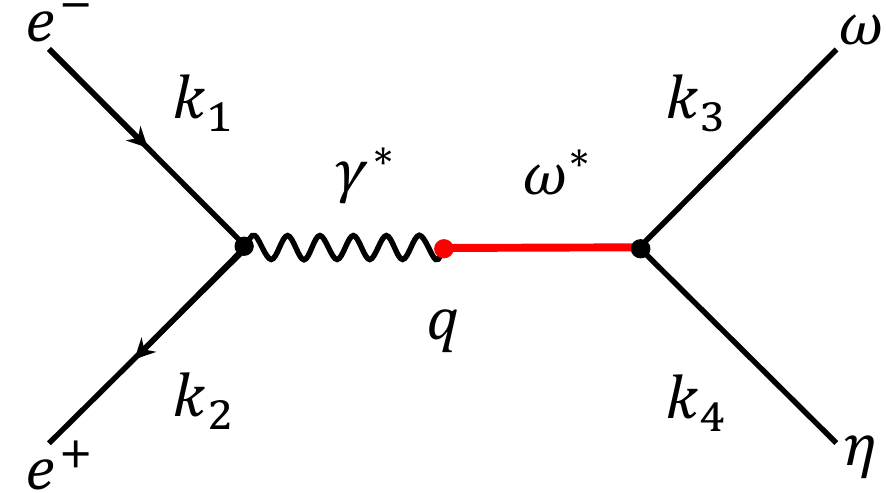}\\
  (a)&(b)
  \end{tabular}
  \caption{The schematic diagrams depicting the reaction $e^+e^-\to\omega\eta$. Here, diagram (a) is the virtual photon directly coupling with final states, while diagram (b) is due to the intermediate state $\omega^*$ contribution, where $\omega^*$ denotes the $\omega(4S)/\omega(3D)$.}\label{Feyomegaeta}
\end{figure}

For the $e^+e^-\to\omega\eta$ process, there exist two possible mechanisms, which have contribution to the cross section. As 
shown in Fig. \ref{Feyomegaeta}, $e^+e^-$ may directly annihilate into virtual photon, which couples with the final state $\omega\eta$. 
The $e^+e^-\to \omega\eta$ process occurs via the intermediate 
$\omega(4S)$ and $\omega(3D)$ for the situation discussed in this work.

When calculating these processes, we adopt effective Lagrangian approach. Here, the effective Lagrangians related to the $e^+e^-\to\omega\eta$ process include \cite{Bauer:1975bv,Bauer:1975bw,Kaymakcalan:1983qq,Lin:1999ad,Oh:2000qr,Zhou:2022ark}
\begin{eqnarray}
\mathcal{L}_{\gamma\mathcal{V}}&=&\frac{-em_{\mathcal{V}}^{2}}{f_{\mathcal{V}}}\mathcal{V}_{\mu}A^{\mu},\\
\mathcal{L}_{\gamma\mathcal{VP}}&=&g_{\gamma\mathcal{VP}}\epsilon_{\mu\nu\alpha\beta}\partial^{\mu}\mathcal{A}^{\nu}\partial^{\alpha}\mathcal{V}^{\beta}\mathcal{P},\\
\mathcal{L}_{\mathcal{VVP}}&=&g_{\mathcal{VVP}}\epsilon_{\mu\nu\alpha\beta}\partial^{\mu}\mathcal{V}^{\nu}\partial^{\alpha}\mathcal{V}^{\beta}\mathcal{P},
\end{eqnarray}
where $\mathcal{V}$ and $\mathcal{P}$ stand for the vector and pseudoscalar meson fields, respectively.

The amplitudes of $e^+e^-\to\omega\eta$ corresponding to the diagrams (a) and (b) in Fig. \ref{Feyomegaeta} can be expressed as
\begin{eqnarray}
\mathcal{M}_{\rm{Dir}}^{\omega\eta}&=&g_{\gamma\omega\eta}\bar{v}(k_1)(ie\gamma_{\mu})u(k_2)\frac{-g^{\mu\nu}}{s}\epsilon_{\rho\nu\alpha\beta}q^\rho k_{3}^{\alpha}\\ \nonumber
&&\times\varepsilon^{*\sigma}(k_3)\mathcal{F}(s),\\
\mathcal{M}^{\omega\eta}&=&g_{\omega^*\omega\eta}\mathcal{M}_{e^+e^-\to\omega^*}^{\rho}\epsilon_{\delta\rho\alpha\beta}q^{\delta} k_3^{\alpha}\varepsilon^{*\beta}(k_3) \label{eq5}, 
\end{eqnarray}
respectively,
where $k_1$, $k_2$, $k_3$ and $k_4$ are four-momentum of $e^+$, $e^-$, $\omega$ and $\eta$, respectively, and $q=k_1+k_2=k_3+k_4$.
$\mathcal{F}(s)=\rm{exp}$$(-b(\sqrt{s}-\sum_f m_f))$ denotes the form factor, where the $b$ is free parameters, which can be determined by fitting experimental data, and $\sum_f m_f$ is the sum for the masses of the final particles. 
The coupling constant $g_{\omega^* \omega \eta}$ stands for $g_{\omega(4S) \omega \eta}$ or $g_{\omega(3D) \omega \eta}$. 
With the values of the branching ratio of $\omega(4S)$/$\omega(3D)$ decay to $\omega \eta$ shown in Fig. \ref{OZIdecays}, the $g_{\omega(4S) \omega \eta}$ and $g_{\omega(3D) \omega \eta}$ are estimated to be $g_{\omega(4S) \omega \eta}=0.5\,\rm{GeV}^{-1}$ or $g_{\omega(3D) \omega \eta}=0.3\,\rm{GeV}^{-1}$.
Furthermore, $\mathcal{M}_{e^+e^-\to\omega^*}^{\rho}$ is the amplitude of $e^+e^-\to \omega^*$
\begin{eqnarray}
\mathcal{M}_{e^+e^-\to\omega^*}^{\rho}=\bar{v}(k_1)(ie\gamma_{\mu})u(k_2)\frac{-g^{\mu\nu}}{s}\frac{-em_{\omega^*}^2}{f_{\omega^*}}\frac{\tilde{g}_{\nu\rho}(p_1)}{s-m_{\omega^*}^2+im_{\omega^*\Gamma_{\omega^*}}}\nonumber
\end{eqnarray}
with $\tilde{g}_{\mu\nu}(q)=-g_{\mu\nu}+q^\mu q^\nu/q^2$, and the resonance parameters of $\omega(4S)$ and $\omega(3D)$ are taken from these values as shown in Fig. \ref{OZIdecays}. 
The parameter $f_{\omega^*}$ can be related to the  dilepton decay width of $\omega^*$ by
\begin{eqnarray}
\Gamma_{\omega^* \to e^+e^-}=\frac{e^4 m_{\omega^*}}{12 \pi f_{\omega^*}^2} \label{dilepton}
\end{eqnarray} 
Here, we adopt the $\Gamma_{\omega(4S) \to e^+e^-}=7 \, \rm{eV}$ and $\Gamma_{\omega(4S) \to e^+e^-}=1.8 \, \rm{eV}$ \cite{Wang:2021gle}, which are estimated according to the zero-point behavior of their radial
wave functions.
Therefore, using the Eq. (\ref{dilepton}), we obtain decay constants $f_{\omega(4S)}=264$ and $f_{\omega(3D)}=532$.

With the above amplitudes, the differential cross sections of $e^+e^-\to\omega\eta$  can be calculated by
\begin{eqnarray}
d\sigma=\frac{1}{32\pi s}\frac{|\vec{k}_{3\rm{cm}}|}{|\vec{k}_{1cm}|}|\overline{\mathcal{M}_{\rm{Total}}^{\omega\eta}|^2}d\rm{cos}\theta,
\end{eqnarray}
where $\theta$ is the scattering angle of an outgoing $\omega$ relative to the direction of the electron beam in the center-of-mass frame,
while $\vec{k}_{1\rm{cm}}$ and $\vec{k}_{3\rm{cm}}$ are the three-momentum of the electron and $\omega$ in the center-of-mass frame, respectively. The overline in $|\overline{\mathcal{M}_{\rm{Total}}^{\omega\eta}}|^2$ indicates the average over the polarization of $e^+e^-$ in the initial states and the sum over the polarization of $\omega\eta$ in the final states.
In addition, $\mathcal{M}_{\rm{Total}}^{\omega\eta}$ is the total amplitude of $e^+e^-\to\omega\eta$, which is expressed as follows
\begin{eqnarray}
\mathcal{M}_{\rm{Total}}^{\omega\eta}=\mathcal{M}_{\rm{Dir}}^{\omega\eta}+\sum_{R}\mathcal{M}_{R}^{\omega\eta}e^{i\phi_R},\label{Eq7}
\end{eqnarray}
where $R$ stands for different intermediate $\omega^*$ meson states, and $\phi_R$ denotes the phase angle of amplitudes of the direct annihilation and the intermediate excited $\omega$ meson contribution.

\subsection{The $e^+e^-\to \omega\pi^0\pi^0$ process}\label{sec2C}

\begin{figure*}[htbp]
  \centering
  \begin{tabular}{ccc}
  \includegraphics[width=150pt]{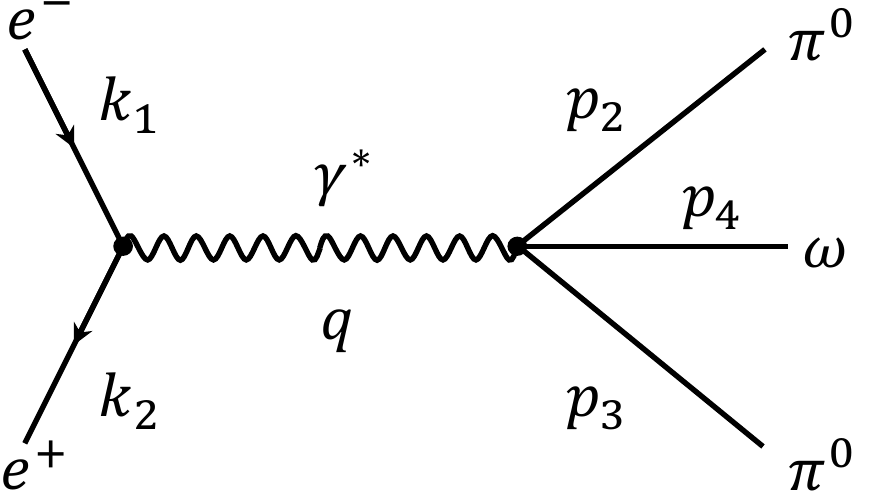}&
  \includegraphics[width=150pt]{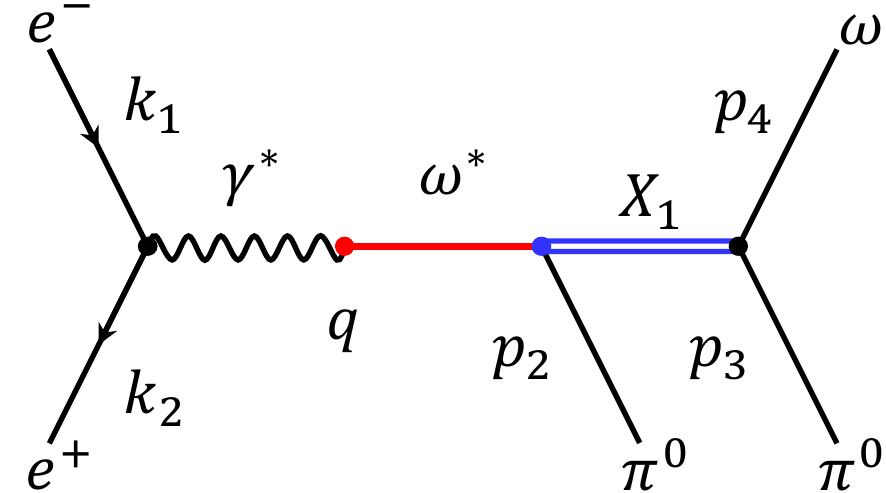}&\includegraphics[width=150pt]{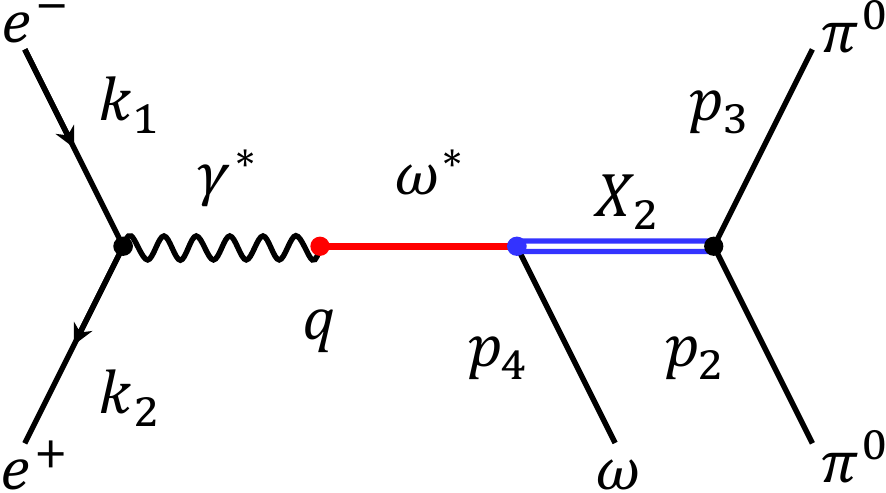}\\
  (a)&(b)&(c)\\
  \end{tabular}
  \caption{Feynman diagrams depicting the mechanisms of the reaction $e^+e^-\to\omega\pi^0\pi^0$.
  Here diagrams (a) corresponds to direct annihilation, and diagrams (b) and (c) present the contributions from the intermediate $\omega^*$ mesons via cascade decays. Here, $\omega^*$ denotes $\omega(4S)/\omega(3D)$, while $X_1$ denotes $\rho$/$\rho(1450)$/$b_1(1235)$ and $X_2$ represents $f_2(1270)$.}\label{Feyomegapipi}
\end{figure*}

As shown in Fig. \ref{Feyomegapipi}, three typical reaction mechanisms for the $e^+e^-\to \omega\pi^0\pi^0$ process are given. Besides continuum contribution, the intermediate states with the $\omega(4S)$ and $\omega(3D)$ may play role in $e^+e^-\to \omega\pi^0\pi^0$, where the  $\omega(4S)$ and $\omega(3D)$ can be coupled to the final state by the cascade processes $e^+e^-\to \omega ^* \to \pi^0 X_1  \to \omega \pi^0\pi^0$ and $e^+e^-\to \omega ^* \to \omega X_2 \to \omega \pi^0\pi^0$. Supported by the study of decay properties of the $\omega(4S)$ and $\omega(3D)$ (see Fig. \ref{OZIdecays}),  
we may determine the concrete particles corresponding to $X_1$
and $X_2$, i.e., 
$X_1$ denotes $\rho$/$\rho(1450)$/$b_1(1235)$, while $X_2$ represents $f_2(1270)$.


The involved effective Lagrangians of $e^+e^-\to \omega\pi^0\pi^0$ are \cite{Chen:2011cj,Huang:2016gdt,Cheng:2016hxi}
\begin{eqnarray}
\mathcal{L}_{\gamma\mathcal{VPP}}&=&-g_{\gamma\mathcal{VPP}}\mathcal{A}^{\mu}\mathcal{V}^{\mu}\mathcal{P}\mathcal{P},\\
\mathcal{L}_{\mathcal{VVP}}&=&g_{\mathcal{VVP}}\epsilon_{\mu\nu\alpha\beta}\partial^{\mu}\mathcal{V}^{\nu}\partial^{\alpha}\mathcal{V}^{\beta}\mathcal{P},\\
\mathcal{L}_{\mathcal{VVT}}&=&g_{\mathcal{VVT}}\mathcal{V}^{\mu}\mathcal{V}^{\nu}\mathcal{T}_{\mu\nu},\\
\mathcal{L}_{\mathcal{TPP}}&=&g_{\mathcal{TPP}}(\mathcal{P}\partial^{\mu}\partial^{\nu}\mathcal{P}+\partial^{\mu}\partial^{\nu}\mathcal{P}\mathcal{P}-2\partial^{\mu}\mathcal{P}\partial^{\nu}\mathcal{P})\mathcal{T}_{\mu\nu},\\
\mathcal{L}_{\mathcal{VP}b_1}&=&ig_{\mathcal{VP}b_1} V_{\mu}\mathcal{P}b_{1}^{\mu},
\end{eqnarray}
where $\mathcal{T}$ and $b_1$ are the tensor and $b_1$ meson fields, respectively.

We may further write out the amplitudes of $e^+e^-\to \omega\pi^0\pi^0$ 
\begin{widetext}
\begin{eqnarray}
\mathcal{M}_{\rm{Dir}}^{\omega\pi^0\pi^0}&=&-g_{\gamma\omega\pi^0\pi^0}\bar{v}(k_1)(ie\gamma_{\mu})u(k_2)\frac{-g^{\mu\nu}}{s}g_{\nu\sigma}\varepsilon^{*\sigma}(p_4)\mathcal{F}(s),\\
\mathcal{M}_{\rho^*}^{\omega\pi^0\pi^0}&=&g_{\omega^*\rho^*\pi^0}g_{\rho^*\omega\pi^0}\mathcal{M}_{e^+e^-\to\omega^*}^{\rho}\left[\epsilon_{\theta\rho\alpha\beta} q^\theta (p_3^{\alpha}+p_4^{\alpha})
\frac{\tilde{g}^{\beta\tau}(p_3+p_4)}{(p_3+p_4)^2-m_{\rho^*}^2+i m_{\rho^*}\Gamma_{\rho^*}}
\epsilon_{\kappa\tau\delta\sigma}(p_3^{\kappa}+p_4^{\kappa}) p_4^\delta \right.\\ \nonumber
&&\left. +\epsilon_{\theta\rho\alpha\beta} q^\theta (p_2^{\alpha}+p_4^{\alpha})
\frac{\tilde{g}^{\beta\tau}(p_2+p_4)}{(p_2+p_4)^2-m_{\rho^*}^2+i m_{\rho^*}\Gamma_{\rho^*}}
\epsilon_{\kappa\tau\delta\sigma}(p_2^{\kappa}+p_4^{\kappa}) p_4^\delta \right]\varepsilon^{*\sigma}(p_4) \label{eq15},\\
\mathcal{M}_{b_1}^{\omega\pi^0\pi^0}&=&-g_{\omega^*b_1\pi^0}g_{b_1\omega\pi^0}\mathcal{M}_{e^+e^-\to\omega^*}^{\rho}\left[\frac{\tilde{g}_{\rho\sigma}(p_3+p_4)}{(p_3+p_4)^2-m_{b_1}^2+i m_{b_1}\Gamma_{b_1}}+\frac{\tilde{g}_{\rho\sigma}(p_2+p_4)}{(p_2+p_4)^2-m_{b_1}^2+i m_{b_1}\Gamma_{b_1}}\right]\varepsilon^{*\sigma}(p_4)\label{eq16},\\
\mathcal{M}_{f_2}^{\omega\pi^0\pi^0}&=&ig_{\omega^*\omega f_2}g_{f_2\pi^0\pi^0}\mathcal{M}_{e^+e^-\to\omega^*}^{\rho}\frac{\tilde{g}_{\rho\sigma\alpha\beta}(p_3+p_3)}{(p_2+p_3)^2-m_{f_2}^2+i m_{f_2}\Gamma_{f_2}}(p_{2}^{\alpha}p_{2}^{\beta}+p_{3}^{\alpha}p_{3}^{\beta}-2p_{2}^{\alpha}p_{3}^{\beta})\varepsilon^{*\sigma}(p_4) \label{eq17},
\end{eqnarray}
\end{widetext}
where $p_2$, $p_3$, and $p_4$ are four momenta of final states $\pi^0$, $\pi^0$ and $\omega$, respectively, and 
$\tilde{g}_{\rho\sigma\alpha\beta}=\frac{1}{2}(\tilde{g}_{\rho\alpha}\tilde{g}_{\sigma\beta}+\tilde{g}_{\rho\beta}\tilde{g}_{\sigma\alpha})-\frac{1}{3}\tilde{g}_{\rho\sigma}\tilde{g}_{\alpha\beta}$. 
The $\rho^*$, $b_1$ and $f_2$ denote $\rho$/$\rho(1450)$, $b_1(1235)$ and $f_2(1270)$, respectively, where their resonance parameters are taken from PDG \cite{ParticleDataGroup:2020ssz}, which are collected in Table \ref{putinparameters}. 
The coupling constants included in Eqs. (\ref{eq15}) - (\ref{eq17}) are calculated from the branching ratios of the corresponding decay modes, which are also collected in Table \ref{putinparameters}.
Here, the coupling constants involved with the  $\omega(4S)$ and $\omega(3D)$ can be calculated from the branching ratios of the corresponding decay modes given in Fig. \ref{OZIdecays}.
The branching ratio of the $\rho(1450)$ decaying to $\omega\pi^0$ is adopted to be $60\%$, which is estimated by the values of $\mathcal{B}(\rho(1450)\to e^+e^-)\times\mathcal{B}(\rho(1450)\to\omega\pi)=3.7\times10^{-6}$ \cite{ParticleDataGroup:2020ssz} and $\mathcal{B}(\rho(1450)\to e^+e^-)=6.2\times10^{-6}$ \cite{Wang:2021gle}, and the branching ratio of the $b_{1}(1235)$ decay into $\omega\pi^0$ is taken as $100\%$ \cite{Chen:2015iqa}.
Besides, the coupling constant $g_{\rho\omega\pi^0}$ is fixed to be $16.0\, \rm{GeV}^{-1}$ as estimated by the QCD sum rules \cite{Lublinsky:1996yf}.

\begin{table*}
  \centering
  \caption{The input parameters in our calculations. Here, these resonance parameters of the states involved in this work are adopted the values provided by PDG \cite{ParticleDataGroup:2020ssz}, the coupling constants are calculated from the branching ratios of the corresponding decay modes.
  The branching ratio of the $\rho(1450)$ decay into $\omega\pi^0$ is obtained by values of $\mathcal{B}(\rho(1450)\to e^+e^-)\times\mathcal{B}(\rho(1450)\to\omega\pi)=3.7\times10^{-6}$ \cite{ParticleDataGroup:2020ssz} and $\mathcal{B}(\rho(1450)\to e^+e^-)=6.2\times10^{-6}$ \cite{Wang:2021gle}, and the branching ratio of the $b_{1}(1235)$ decay into $\omega\pi^0$ is taken as $100\%$ estimated by the QPC model \cite{Chen:2015iqa}.}\label{putinparameters}
  \begin{tabular}{ccccccccccccccc}
  \toprule[1pt]
  \midrule[1pt]
  Parameters & $\quad$ & Values & $\quad$ & Parameters & $\quad$ & Values & $\quad$ & Parameters & $\quad$ & Values & $\quad$ & Parameters & $\quad$ & Values\\
  \midrule[1pt]
  $m_{\rho(1450)}$ & $\quad$ & 1.465 GeV & $\quad$ & $\Gamma_{\rho(1450)}$ & $\quad$ & 0.400 GeV & $\quad$ & $\mathcal{B}(\rho(1450)\to\omega \pi^0)$ & $\quad$ & 60\% & $\quad$ & $g_{\rho(1450)\omega\pi^0}$ & $\quad$ & 8.2 $\rm{GeV}^{-1}$\\
  $m_{b_{1}(1235)}$ & $\quad$ & 1.230 GeV & $\quad$ & $\Gamma_{b_{1}(1235)}$ & $\quad$ & 0.142 GeV & $\quad$ & $\mathcal{B}({b_{1}(1235)}\to\omega \pi^0)$ & $\quad$ & 100\% & $\quad$ & $g_{b_{1}(1235)\omega\pi^0}$ & $\quad$ & 3.8 $\rm{GeV}$\\
  $m_{f_{2}(1270)}$ & $\quad$ & 1.275 GeV & $\quad$ & $\Gamma_{f_{2}(1270)}$ & $\quad$ & 0.187 GeV & $\quad$ & $\mathcal{B}({f_{2}(1270)}\to\pi \pi)$ & $\quad$ & 84.2\% & $\quad$ & $g_{f_{2}(1270)\pi\pi}$ & $\quad$ & 3.3 $\rm{GeV}^{-2}$\\
  $g_{\omega(4S)\rho\pi^0}$ & $\quad$ & 0.5 $\rm{GeV}^{-1}$ & $\quad$ & $g_{\omega(4S)\rho(1450)\pi^0}$ & $\quad$ & 1.4 $\rm{GeV}^{-1}$ & $\quad$ & $g_{\omega(4S)b_{1}(1235)\pi^0}$ & $\quad$ & 0.8 GeV & $\quad$ & $g_{\omega(4S)\omega f_{2}(1235)}$ & $\quad$ & 0.5 \\
  $g_{\omega(3D)\rho\pi^0}$ & $\quad$ & 0.3 $\rm{GeV}^{-1}$ & $\quad$ & $g_{\omega(3D)\rho(1450)\pi^0}$ & $\quad$ & 0.6 $\rm{GeV}^{-1}$ & $\quad$ & $g_{\omega(3D)b_{1}(1235)\pi^0}$ & $\quad$ & 1.3 GeV & $\quad$ & $g_{\omega(3D)\omega f_{2}(1235)}$ & $\quad$ & 0.5 \\
\midrule[1pt]
\bottomrule[1pt]
\end{tabular}
\end{table*}

The cross section of $e^+e^-\to \omega\pi^0\pi^0$ is
\begin{eqnarray}
d\sigma=\frac{1}{32(2\pi)^5 \sqrt{s}\sqrt{k_1 k_2}}|\overline{\mathcal{M}_{\rm{Total}}^{\omega\pi^0\pi^0}|^2}|\vec{p}_2||\vec{p}_3^*|d\Omega_2 d\Omega_3^* dm_{34},
\end{eqnarray}
where $\vec{p_2}$ ($\Omega_3$) is the the three-momentum (solid angle) of $\pi^0$ in the center-of-mass frame, $\vec{p}_3^*$ ($\Omega_3^*$) stands for the three-momentum (solid angle) of the $\omega$ in the rest frame of $\omega$ and $\pi^0$, and $m_{34}$ is the invariant mass of the
$\omega$ and $\pi^0$ system. Thus, the total amplitude of $\mathcal{M}_{\rm{Total}}^{\omega\pi^0\pi^0}$ is
\begin{eqnarray}
\mathcal{M}_{\rm{Total}}^{\omega\pi^0\pi^0}=\mathcal{M}_{\rm{Dir}}^{\omega\pi^0\pi^0}+\sum_{R,X}\mathcal{M}_{RX}^{\omega\pi^0\pi^0}e^{i\phi_{RX}},\label{Eq18}
\end{eqnarray}
where $R$ and $X$ denote the $\omega(4S)$/$\omega(3D)$ and the intermediate $X_1/X_2$ state, respectively, and $\phi_{RX}$ is the phase angle among different amplitudes.

\section{numerical results}\label{sec3}

\begin{figure}[!htbp]
  \centering
  \begin{tabular}{c}
  \includegraphics[width=220pt]{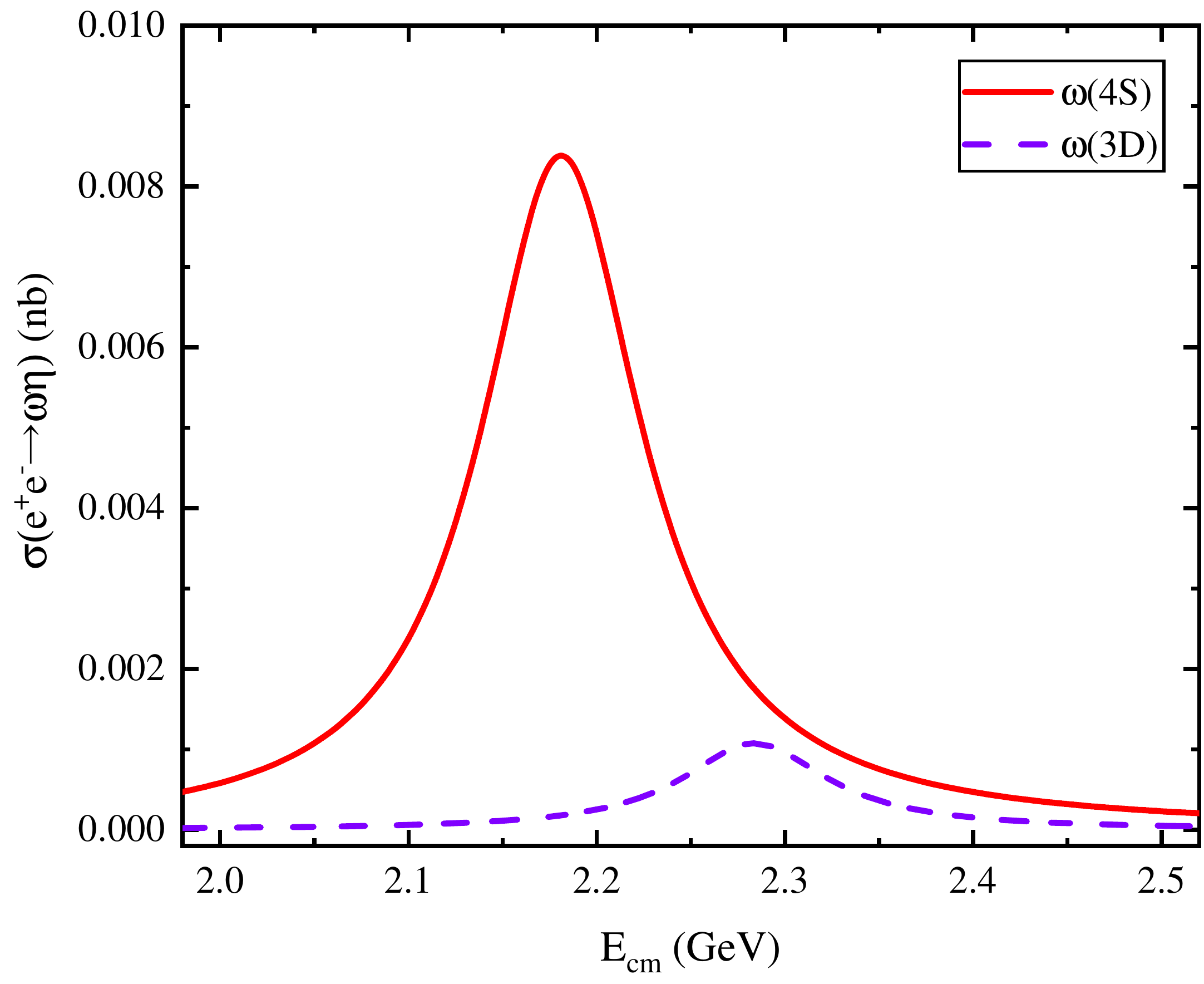}\\
  \end{tabular}
  \caption{The calculated cross sections of $e^+e^-\to\omega(4S)\to\omega\eta$ and $e^+e^-\to\omega(3D)\to\omega\eta$.}\label{oecomp}
\end{figure}

\begin{figure}[!htbp]
  \centering
  \begin{tabular}{c}
  \includegraphics[width=220pt]{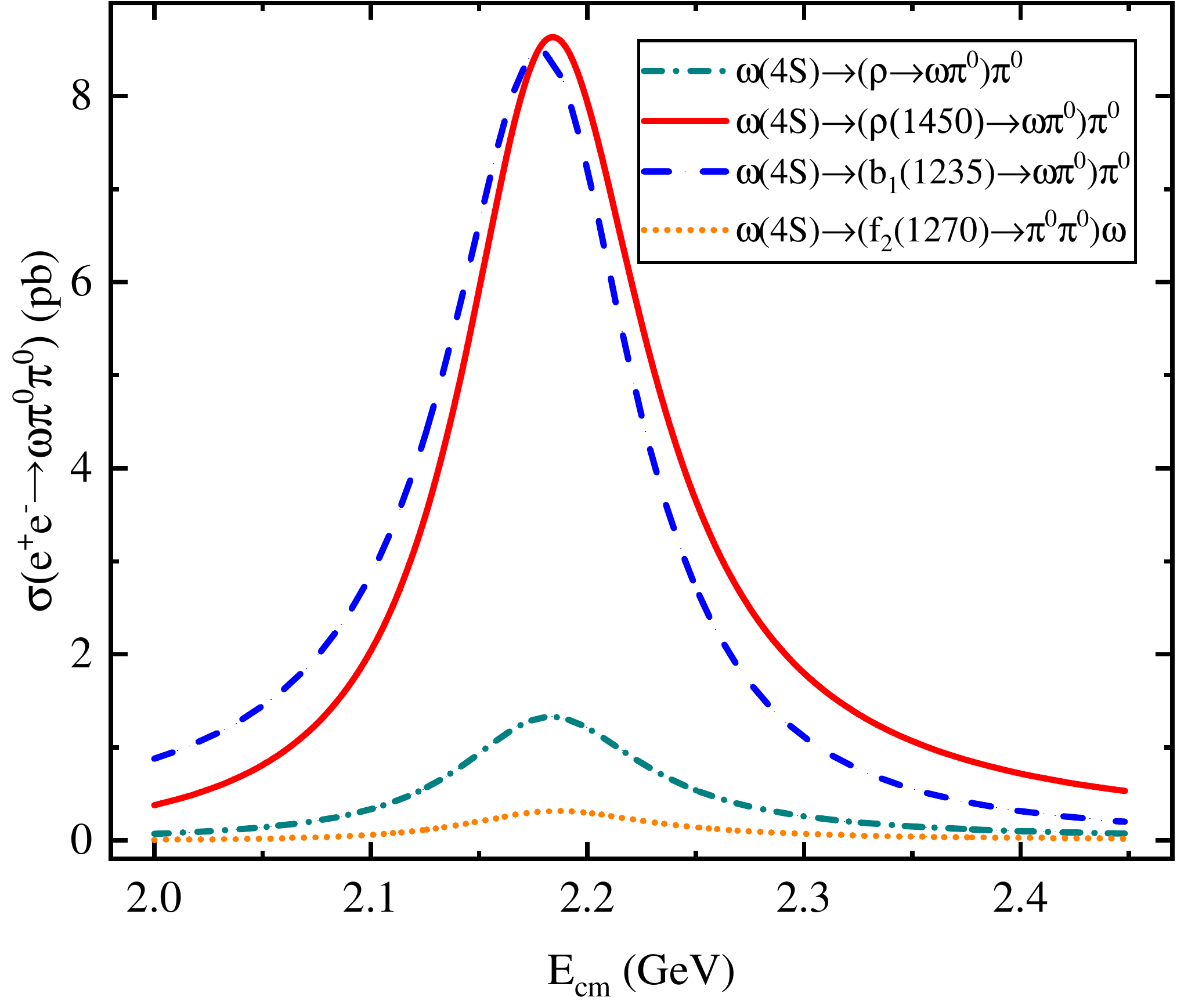}\\
  \includegraphics[width=220pt]{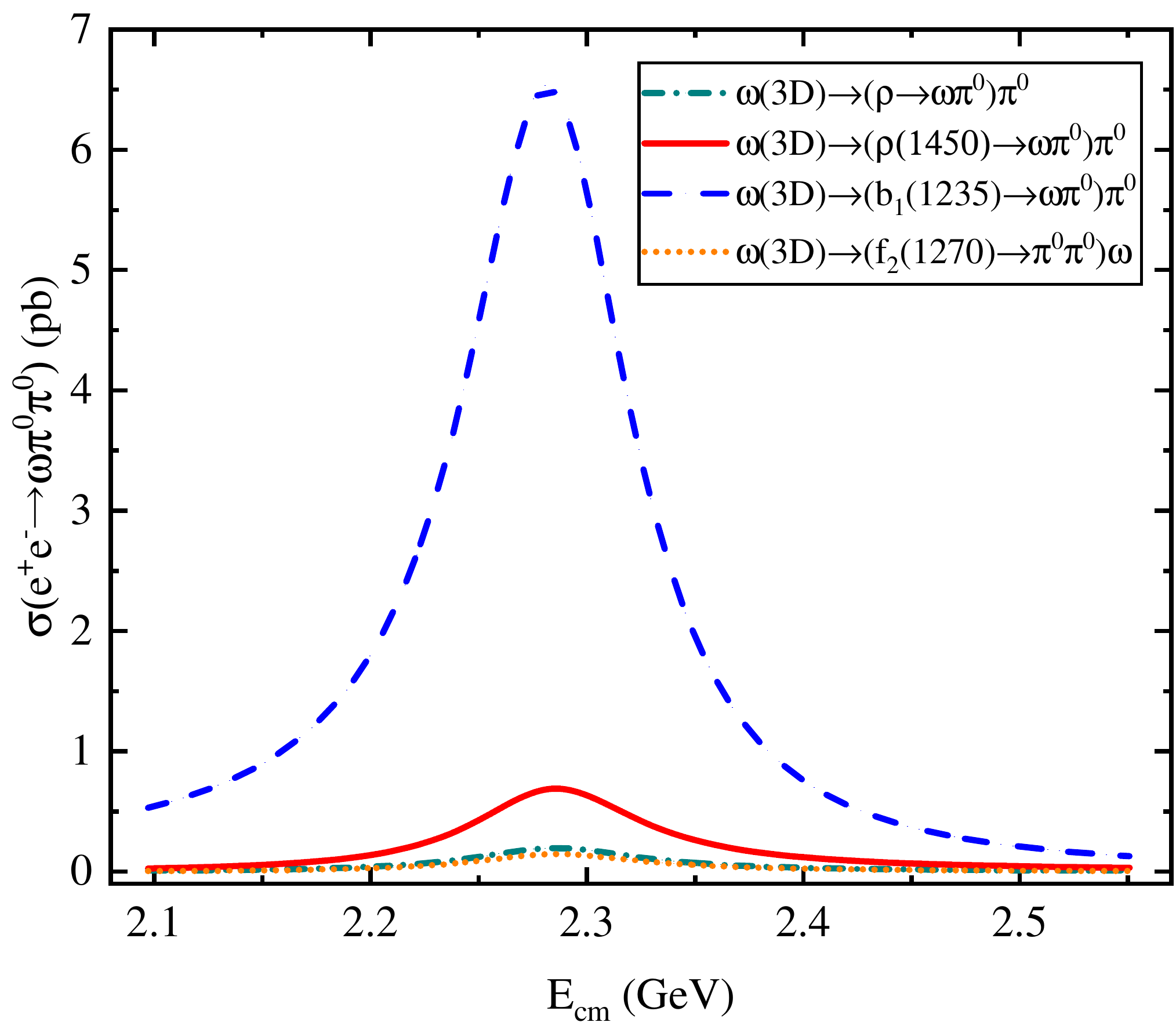}\\
  \end{tabular}
  \caption{The calculated cross sections of $e^+e^-\to\omega(4S)\to\omega\pi^0\pi^0$ and $e^+e^-\to\omega(3D)\to\omega\pi^0\pi^0$ through different cascade processes.}\label{oppcomp}
\end{figure}

In this section, we perform a fit on the experimental data of the Bonn cross sections of $e^+e^-\to\omega\eta$ and  $\omega\pi^0\pi^0$ measured by the BESIII Collaboration \cite{BESIII:2020xmw,BESIII:2021uni} to decipher the contributions of excited $\omega$ meson states, which can help us understand the nature of the enhancement structures around 2.2 GeV observed in these two processes.
Based on the theoretical studies of the spectrum and decay properties of higher $\omega$ mesons near 2.2 GeV shown in Sec. \ref{sec2}, we can directly estimate the cross section sizes of $e^+e^-\to \omega(4S) /\omega(3D) \to \omega\eta$ and  $\omega\pi^0\pi^0$, where the corresponding results are shown in Figs. \ref{oecomp} and \ref{oppcomp}, respectively.

As given in Fig. \ref{oecomp}, the cross section of $e^+e^-\to \omega \eta$ occurring through the intermediate state $\omega(4S)$ is significantly larger than that occurring through the $\omega(3D)$, but they are still in the same order of magnitude.
In Fig. \ref{oppcomp}, we can find that the dominant resonance contributions to $e^+e^-\to\omega\pi^0\pi^0$ are from the  cascade processes $\omega(4S)\to (\rho(1450) \to\omega \pi^0) \pi^0$, $\omega(4S)\to (b_1(1235) \to\omega \pi^0) \pi^0$ and $\omega(3D)\to (b_1(1235) \to\omega \pi^0) \pi^0$. Here, the contributions of these cascade processes $\omega(4S)\to (\rho \to\omega \pi^0) \pi^0$ and $\omega(3D)\to (\rho(1450) \to\omega \pi^0) \pi^0$ are secondary, while the contributions of these cascade processes $\omega(4S)\to (f_2(1270) \to\pi^0 \pi^0)\omega$, $\omega(3D)\to (\rho \to\omega \pi^0) \pi^0$ and $\omega(3D)\to (f_2(1270) \to\pi^0 \pi^0)\omega$ are minor, which are one or two orders of magnitude smaller than the others.
Besides, although the $\rho\pi$ is one of the dominant decay channels of the $\omega(4S)$ and the $\rho$ can be strongly coupled to $\omega\pi$, the contribution of the $\omega(4S)\to (\rho \to\omega \pi^0) \pi^0$ cascade process to $e^+e^-\to\omega\pi^0\pi^0$ is suppressed by an off-shell intermediate $\rho$ state.


After the above preparation, we can fix the theoretical contributions of the $\omega(4S)$ and $\omega(3D)$ to fit the experimental data of the Born cross sections of $e^+e^-\to\omega\eta$ and $e^+e^-\to \omega\pi^0\pi^0$.
In the next fitting procedure, the free parameters only include relative phase angles of the different amplitudes, $g_{\gamma\omega\eta}$, $g_{\gamma\omega\pi^0\pi^0}$, $b_e$ and $b_p$ in the direct production amplitudes.

\begin{table}[!htb]
  \centering
  \caption{The parameters obtained by fitting the experimental data of the Born cross sections of $e^+e^-\to\omega\eta$  \cite{BESIII:2020xmw}, and the $\chi^2/\rm{n.d.f}$ value is 1.59 for this fitting.}\label{reslutsToe}
  \begin{tabular}{ccccc}
  \toprule[1pt]
  \midrule[1pt]
  Parameters $\quad$ &  Values $\quad$ & Parameters $\quad$ & Values\\
  \midrule[1pt]
  $g_{\gamma\omega\eta}$ ($\rm{GeV}^{-1}$)  $\quad$  & $0.053\pm0.001$  $\quad$ & $b_{\omega\eta}$ ($\rm{GeV}^{-1}$)  $\quad$  & $1.63\pm0.05$\\
  $\phi_{\omega(4S)}$ (rad)  $\quad$ & $2.79\pm0.15$  $\quad$ &
  $\phi_{\omega(3D)}$ (rad) $\quad$ &  $6.17\pm0.47$ \\
 \midrule[1pt]
\bottomrule[1pt]
\end{tabular}
\end{table}

\begin{figure}[!htbp]
  \centering
  \begin{tabular}{c}
  \includegraphics[width=240pt]{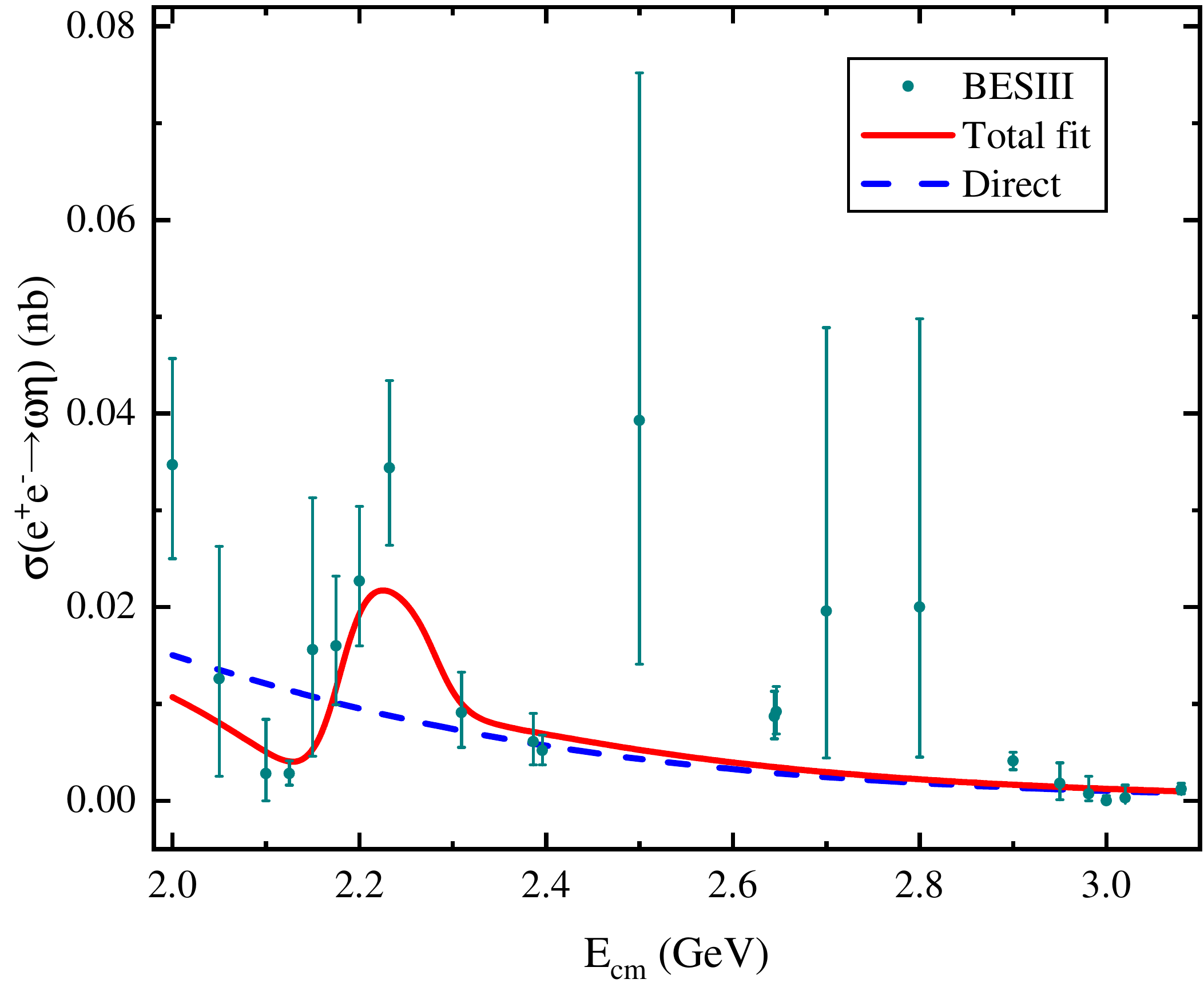}
  \end{tabular}
  \caption{The fitted result of the experimental data of the Born cross sections of $e^+e^-\to\omega \eta$ \cite{BESIII:2020xmw}.}\label{reslutsFoe}
\end{figure}

The fitted parameters for the $e^+e^-\to\omega\eta$ reaction are collected in Table \ref{reslutsToe}, and the obtained $\chi^2/\rm{n.d.f}$ value is 1.59.
With the central values of the fitted parameters, we can obtain the curve of the total cross sections of $e^+e^-\to \omega\eta$, which is presented in Fig. \ref{reslutsFoe} by a black solid line, where the fitted line shape can well reproduce the experimental data.
Obviously, the $\omega(4S)$ plays a dominant role in the $e^+e^-\to \omega\eta$ process compared to the $\omega(3D)$, especially for reproducing enhancement structure around 2.2 GeV.
Therefore, the line shape of the total cross section of $e^+e^-\to \omega \eta$  mainly reflects the characteristics of the  $\omega(4S)$. Thus, we understand why the resonance parameter reported by BESIII in the process of $e^+e^-\to \omega \eta$ is comparable with the mass and width of the $\omega(4S)$ predicted in Ref. \cite{Wang:2021gle}.

\begin{table}[!htb]
  \centering
  \caption{The parameters obtained by fitting the experimental data of the Born cross sections of $e^+e^-\to\omega\pi^0\pi^0$ \cite{BESIII:2021uni}, and the $\chi^2/\rm{n.d.f}$ value is 1.79 for this fitting.}\label{reslutsTopp}
  \begin{tabular}{cccc}
  \toprule[1pt]
  \midrule[1pt]
   Parameters $\,$ & Values $\,$  & Parameters $\,$ & Values\\
  \midrule[1pt]
$g_{\gamma\omega\pi^0\pi^0}$ $\,$ & $4.84\pm0.02$ $\,$ & $b_{\omega\pi^0\pi^0}$ ($\rm{GeV}^{-1}$)  $\,$ & $1.12\pm0.01$\\
  $\phi_{\omega(4S)\rho}$ (rad) $\,$ &  $1.82\pm0.55$ $\,$ & $\phi_{\omega(4S)\rho(1450)}$ (rad) $\,$ & $0.54\pm0.23$\\
  $\phi_{\omega(4S)b_1(1235)}$ (rad) $\,$ & $4.81\pm0.03$ $\,$ &
  $\phi_{\omega(3D)\rho(1450)}$ (rad) $\,$ & $5.34\pm0.72$\\
  $\phi_{\omega(3D)b_1(1235)}$ (rad) $\,$ & $1.79\pm0.05$ $\,$ & 
  $\quad$ &  $\quad$ \\
 \midrule[1pt]
\bottomrule[1pt]
\end{tabular}
\end{table}

\begin{figure}[!htbp]
  \centering
  \begin{tabular}{c}
   \includegraphics[width=240pt]{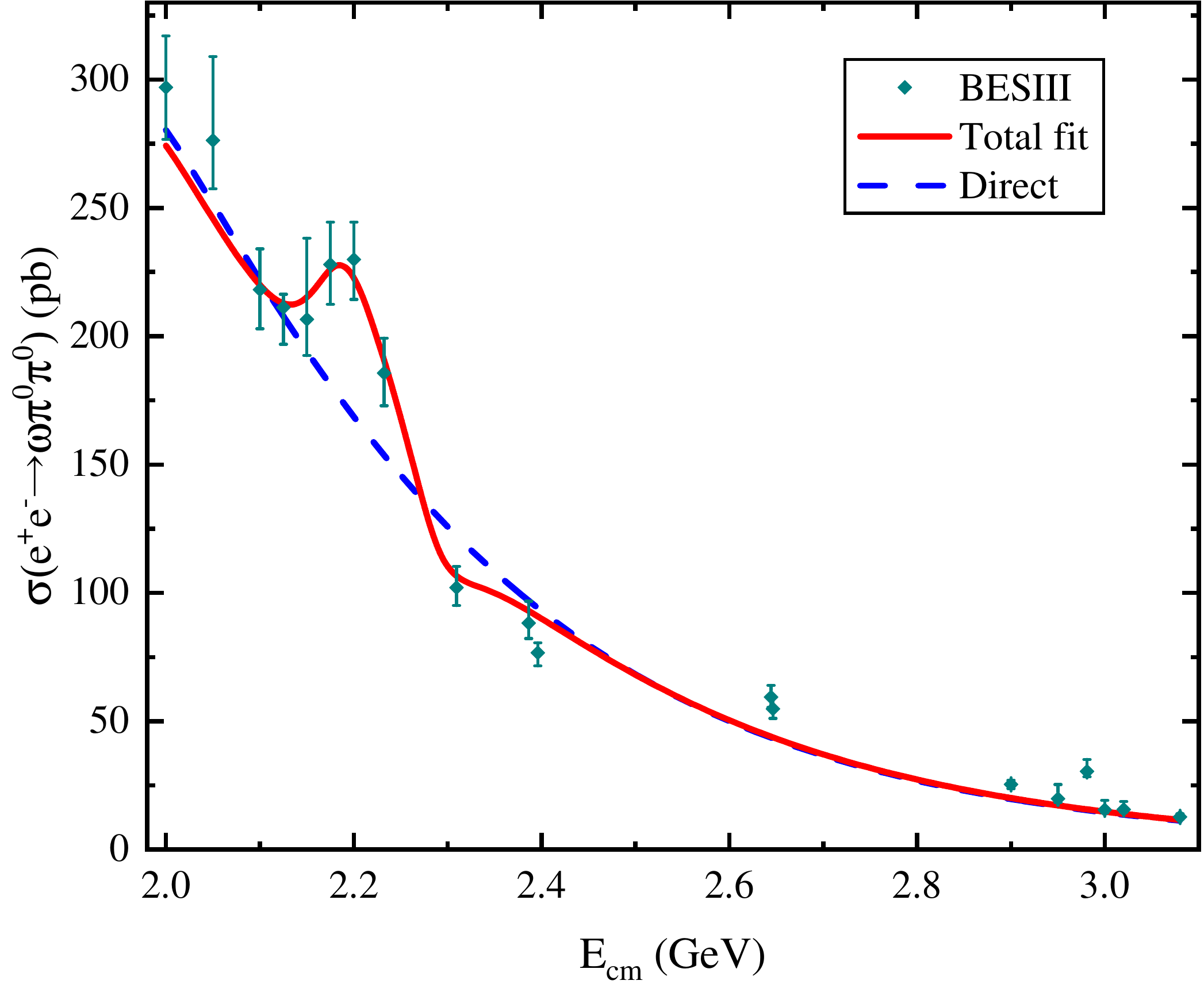}
  \end{tabular}
  \caption{The fitted result of the experimental data of the Born cross sections of $e^+e^-\to\omega \pi^0\pi^0$ \cite{BESIII:2021uni}.}\label{reslutsFopp}
\end{figure}

Both the $\omega(4S)$ and $\omega(3D)$ can contribute to the measured $e^+e^-\to\omega\pi^0\pi^0$ process through multiple cascade processes as discussed above.
In order to reduce the number of free fitting parameters, these contributions from the cascade processes $\omega(4S)\to (f_2(1270) \to\pi^0 \pi^0)\omega$, $\omega(3D)\to (\rho \to\omega \pi^0) \pi^0$ and $\omega(3D)\to (f_2(1270) \to\pi^0 \pi^0)\omega$ to $e^+e^-\to \omega \pi^0\pi^0$ are not considered when fitting the experimental data of the Born cross section of $e^+e^-\to\omega\pi^0\pi^0$, since these contributions are one or two orders of magnitude smaller than the others.
The fitted results for the experimental data of $e^+e^-\to\omega\pi^0\pi^0$ are presented in Table \ref{reslutsTopp} and Fig. \ref{reslutsFopp}, where the obtained $\chi^2/\rm{n.d.f}$ value is 1.79. Here,
it can be seen that the enhancement structure observed in $e^+e^-\to\omega\pi^0\pi^0$ near 2.2 GeV can also be reproduced well. 
The interference between different cascade processes is crucial to reproduce the experimental data, which results in an enhancement structure sandwiched by the mass positions of the $\omega(4S)$ and $\omega(3D)$ in the total cross section of the $e^+e^-\to \omega\pi^0\pi^0$ process. 
Through the above analysis, we may draw a conclusion that the enhancement structure observed in $e^+e^-\to\omega\pi^0\pi^0$ near 2.2 GeV can be due to total contribution from the $\omega(4S)$ and $\omega(3D)$.

Based on the above analysis of the cross sections of the $e^+e^-\to\omega\eta$ and $e^+e^-\to\omega\pi^0\pi^0$ processes supported by the theoretical spectroscopy of higher $\omega$ mesonic states,
we find that the enhancement structures near 2.2 GeV reported in $e^+e^-\to \omega\eta$ \cite{BESIII:2020xmw} and $e^+e^-\to \omega\pi^0\pi^0$ \cite{BESIII:2021uni} can be reproduced well by introducing the contributions of the $\omega(4S)$ and $\omega(3D)$.
Since the difference in the relative contribution of the $\omega(4S)$ and $\omega(3D)$ in $e^+e^-\to \omega\eta$ and $e^+e^-\to \omega\pi^0\pi^0$ as well as the interference effect, it naturally explains the difference of the resonance parameters of the enhancement structures around 2.2 GeV reported in these two processes \cite{BESIII:2020xmw,BESIII:2021uni}.
Finally, we can see that the experimental data of $e^+e^-\to \omega\eta$ and $e^+e^-\to \omega\pi^0\pi^0$ 
show the existence of the $\omega(4S)$ and $\omega(3D)$, which is consistent with 
former theoretical predictions for the $\omega(4S)$ and $\omega(3D)$ by an unquenched potential model \cite{Wang:2021gle}. It is obvious that the present study provides valuable information to construct the $\omega$ meson family.

\section{Discussion and conclusion}\label{sec4}

Although more and more new hadronic states were observed in the past decades \cite{Chen:2016qju,Liu:2019zoy,Olsen:2017bmm,Brambilla:2019esw}, light flavor meson family is far from being established. How to construct light flavor meson family with the reported experimental phenomena becomes an intriguing research topic \cite{Pang:2019ovr,Wang:2021gle,Wang:2012wa,Zhou:2022ark,He:2013ttg,Pang:2019ttv,Wang:2017iai,Chen:2020xho,Wang:2019qyy,Wang:2020kte,Wang:2020due,Wang:2021abg,Li:2021qgz,Guo:2022xqu}. With data accumulation of $e^+e^-$ collision at $\sqrt{s}\sim 2$ GeV, light flavor vector enhancement structures around 2.2 GeV were reported in $e^+e^-\to \omega\eta$ \cite{BESIII:2020xmw} and $e^+e^-\to \omega\pi^0\pi^0$ \cite{BESIII:2021uni}, which can be related to higher $\omega$ mesonic states. 

In this work, guided by theoretical knowledge of $\omega$ mesonic spectroscopy \cite{Wang:2021gle},
we introduce the $\omega(4S)$ and $\omega(3D)$ to depict the behavior of enhancement structures observed in $e^+e^-\to \omega\eta$ \cite{BESIII:2020xmw} and $e^+e^-\to \omega\pi^0\pi^0$ \cite{BESIII:2021uni}, and find that these two structures around 2.2 GeV are resulted from the interference of the $\omega(4S)$ and $\omega(3D)$ signals. By this way, the puzzling difference of the resonance parameter of these vector enhancement structures can be explained well. 
We should indicate that the present work is the first step when establishing higher $\omega$ mesonic states, which is one part of whole constructing light flavor hadronic spectroscopy. More theoretical and experimental efforts are encouraged.

In the future, BESIII, as main force of exploring light flavor hadron, has good chance to carry out further experimental exploration to identify the $\omega(4S)$ and $\omega(3D)$ states through high precision data of $e^+e^-\to \omega\eta$ and $e^+e^-\to \omega\pi^0\pi^0$.

\section*{ACKNOWLEDGEMENTS}

This work is supported by the China National Funds for Distinguished Young Scientists under Grant No. 11825503, National Key Research and Development Program of China under Contract No. 2020YFA0406400, the 111 Project under Grant No. B20063, the National Natural Science Foundation of China under Grant No. 12047501, the Fundamental Research Funds for the Central
Universities, and Science and Technology Department of Qinghai Province Project No. 2020-ZJ-728.

\end{document}